# ULTIMATE INTERNETS

P. A. Zizzi

Dipartimento di Astronomia dell'Università di Padova
Vicolo dell' Osservatorio, 2
35122 Padova, Italy
zizzi@pd.astro.it

## Abstract

In a previous paper (gr-qc/0103002), the inflationary universe was described as a quantum growing network (QGN).
Here, we propose our view of the QGN as the "ultimate Internet", as it saturates the quantum limits to computation. Also, we enlight some features of the QGN which are related to: i) the problem of causality at the Planck scale, ii) the quantum computational aspects of spacetime foam and decoherence, iii) the cosmological constant problem, iv) the "information loss" puzzle.
The resulting picture is a self-organizing system of ultimate Internet-universes.

# 1. Introduction

It seems that quantum spacetime has a lot to share with quantum information, quantum computation, and quantum communication.
In fact, in [1] we showed that the holographic principle [2] is strictly related to quantum information encoded in quantum spacetime. However, any computational system, must obey the quantum limits to computation, given by the Margolous/Levitin theorem [3]. Thus, quantum computing spacetime, should obey these bounds as well. Indeed, quantum spacetime saturates these bounds, as it will be illustrated in the following.
Recently, Seth Lloyd [4] suggested that black holes can be regarded as ultimate computers, as they saturate the quantum limits to computation. Moreover, Ng showed [5] that spacetime foam [6] provides the quantum limits to computation.
Finally, in a recent paper [7], we described the early inflationary universe as a quantum growing network (QGN), which saturates the quantum limits to computation at each time step. Then the QGN (which can be thought as the first toy model for a future quantum WWW), is in fact a "ultimate Internet". In summary, quantum spacetime itself provides the ultimate tools for quantum computation and quantum communication.
The QGN graph is made of two parts: the connected part which is the ensemble of the connecting links and the nodes, and the disconnected part, which is made of free links outgoing from the nodes. The connecting links are virtual black holes, quantum fluctuations of the vacuum, which carry virtual information (because of the inverse relation between quantum information and the quantized cosmological constant). The nodes are quantum fluctuations of the metric, which operate like quantum logic gates, and transform the virtual information into available information. The free links are the qubits. The ensemble of virtual black holes and quantum fluctuations of the metric is what is called here "quantum foam". Then, the ultimate Internet appears to be an artifact of spacetime foam.
Interestingly, the QGN leads to a new concept, the causality degree (CD), which somehow "measures" the causal structure of the net during its growth. The result is that at the very Planck scale, the causality degree is zero. Hovever, as the causal structure is encoded in the program and emerges during the growth, we associate the zero causality degree to what we call "proto-causality" at the Planck scale.
If it is true that the QGN saturates the quantum limits to computation, it does that at each time step, because of its growth. This is a different case from that of the ultimate computer, where in principle the imput power (which limits the speed of computation) is known once for all. In our case, it is not possible to know, for example, whether there is a maximum amount of quantum information processed by the QGN. The knowledge of that would be necessary to state *a priori* when the QGN decohered, giving rise to a clssical universe (we racall, moreover, that the decoherence of the QGN would coincide with the end of inflation [7]). Nevertheless, there are a few hints (considerations on the actual entropy of our universe) which allow us to identify the speed of computation and memory space at the moment of decoherence. At this point, one could raise objections concerning decoherence of an isolated system. However, as we said, the QGN can be considered as made od two subsystems: the connected part, which is the quantum foam, and the disconnected part which is the quantum state of N qubits. We believe that the decoherence is due to the quantum foam which plays the role of the environment. But, as the quantum foam is part of the QGN, the QGN as a whole undergoes self-decoherence.

It should be noticed that, if the quantum state of N qubits did not decohere at some earlier time, the present amount of entropy would be huge.

In this model, the theoretical value of the cosmological constant is in agreement with the Type Ia SN observation data [8]. However, the calculations leading to this result, assume that the connecting links of the QGN (quantum fluctuations of the vacuum) are still active today. Also, the nodes (the quantum fluctuations of the metric) are still present, although they are endowed with a very low mean energy. How reconciling this facts with decoherence? A possible answer would be that after decoherence, only the connected part of the QGN (actually a lattice-like structure) survived. This is the discrete substratum of our smooth spacetime manifold, which becomes apparent at the Planck resolution.

Finally, the QGN is related to the "information loss" puzzle [9] in the following way. If the remnant of a black hole which evaporated, is a Planckian black hole, it cannot store all the information collapsed in the original black hole, but only one qubit. This fact is due to the quantum limits to computation, and can also be explained by the holographic principle. But the remnant gives rise to a QGN (actually to a new universe) where the maximum number of qubits available before decoherence, is equal to the number of bits "lost" in the older universe. The resulting scenario is a self-organizing system of universes, each universe beeing a ultimate Internet at its early stage.

For the previous arguments, the new universes are produced in such a way that their maximum entropy is reduced, and their evolution is made possible.

In Section 2, we give a short review of the QGN model. In Section 3, we introduce the concept of causality degree, and we show that at the Planck scale the causality degree is zero. In Section 4, we discuss some computational aspects of spacetime foam and illustrate its relation with the QGN. In Section 5, we show that the QGN saturates the quantum limits to computation. In Section 6, we illustrate the relation between the QGN and the present value of the cosmological constant. In Section 7, we explain the reason why we believe that the QGN self-decohered at the end of inflation. In Section 8, we illustrate some relations between the QGN and the "information loss" puzzle in black holes evaporation. Section 9, is devoted to some concluding remarks.

## 2. The quantum growing network (QGN): a brief review

In a recent paper [7], the early inflationary universe was described as a quantum growing network (QGN). The speed up of growth of the network (inflation) is due to the presence of virtual qubits in the vacuum state of the quantum memory register. Virtual quantum information is created by quantum vacuum fluctuations, because of the inverse relation [7] between the quantized cosmological constant [10], and quantum information I:

1) $\quad \Lambda_n = \frac{1}{I \, l_P^2}$

where: n=0,1,2,3... and the quantum information I is [1]:

2) $\quad I \equiv N = (n+1)^2$

Eq. 1) leads to:

3) $\quad \Delta I = \frac{-\Delta \Lambda}{l_P^2 \Lambda^2} = 2n + 3$

That is, at each time step $t_n$, there are 2n+3 extra bits (virtual states) in the vacuum state of the quantum memory register, where:

4) $\quad t_n = (n+1)t_P$

is the quantized time, and $t_P \cong 10^{-43}$ sec. is the Planck time. The 2n+3 virtual states occurring at time $t_n$, are operated on by a quantum logic gate $U_n$ [7] at time $t_{n'} = t_{n+1}$, where:

5) $\quad U_n = \prod_{j=1}^{v=2n+3} Had(j)$

and Had (j) is the Hadamard gate operating on bit j. Then, the virtual states, are transformed into 2n'+1 qubits at time $t_{n'}$.

This results into a quantum growing network, where the nodes are the quantum logic gates, the connecting links are the virtual states, and the free links are the qubits. At each time step, the total number of qubits (free links) is $N = (n+1)^2$.

The rules of the growing quantum network that we consider, are resumed below.

At the starting time (the unphysical time $t_{-1} = 0$), there is one node, call it **-1**. At each time step $t_n$, a new node is added, which links to the youngest and the oldest nodes, and also carries 2n+1 free links. Thus, at the Planck time $t_0 = t_P$, the new node **0** is added, which links to node **-1** and carries one free link. At time $t_1 = 2t_P$, the new node **1** is added, which links to nodes **-1** and **0**, and carries three free links. At time $t_2 = 3t_P$, the new node **2** is added, which links to nodes **-1** and **1**, and carries five free links. At time $t_3 = 4t_P$, the new node **3** is added, which links to nodes **-1** and **2**, and carries seven free links, and so on.

In general, at time $t_n$, there are:

1) $n+2$ nodes, but only n+1 of them are active, in the sense that they have outgoing free links (node **-1** has no outgoing free links).
2) $N = (n+1)^2$ free links coming out from n+1 active nodes
3) 2n+1 links connecting pairs of nodes
4) n loops.

The N free links are qubits (available quantum information), the 2n+1 connecting links are virtual states, carrying information along loops, the n+1 active nodes are quantum logic gates operating on virtual states and transforming them into qubits. In fact, notice that the number of free outgoing links at node **n** is 2n+1, which is also the number of virtual states (connecting links) in the loops from node **-1** to node **n**.

The connected part of the network is the most similar to a scale-free growing network, [11] like the World Wide Web, where free links are absent. However, the connected part is deterministic, in the sense that it follows some precise rules, as a lattice. What is missing here, is one of the two fundamental features of scale-free networks, which is preferential attachment. The free links, however, destroy the structure of a regular lattice, as the configuration of free links changes at each time step.

We cannot figure out yet the mathematical model of a quantum WWW, but it will be a quantum growing network, and the QGN described above, seems to be a good starting point. Moreover, as it was shown in [7], the QGN saturates the quantum limits to computation [3] [4] [5], thus it can be viewed as the ultimate Internet. The ultimate computer [4] beeing a black hole, it would not be surprising that the ultimate Internet is quantum spacetime itself.

## 3. Causality degree

Let us consider the quantum growing network discussed above. It is a discrete system which, at time $t_n$, has $N = (n+1)^2$ degrees of freedom. This discrete system can in fact be interpreted as a discrete spacetime with N "events" (the N qubits). It follows that the first "event" has caused N-1 "events". This leads us to revisite the concept of causality in this particular context.

Let us define the (relative) causality degree (CD) at time $t_n$ with respect to time $t_m$ (with n>m) as the increase of quantum information $\Delta I$:

6)  $(CD)_{n,m} = \Delta I = N - M = n^2 - m^2 + 2(n - m)$

(In particular, for m=n+1, we get: $(CD)_{n,n+1} = 2n + 3$).

The (absolute) causality degree at time $t_n$ is then defined with respect to the Planck time (m=0):

7)  $(CD)_n = n^2 + 2n = N - 1$

From eq. 7, we see, for example, that at time $t_1$ the causality degree is 3. At time $t_1$ in fact, the discrete system has 4 degrees of freedom (or four events). The event at time $t_0$ has caused 3 new events.

Moreover, from eq. 7, we find that the absolute causality degree at the Planck time (n=0) is zero:

8)  $(CD)_0 = 0$

This result becomes quite understandable when the quantum fluctuations of the metric are taken into account.

In fact, the expression [10] of the quantum fluctuations of the metric at time $t_n$ is:

9)  $\Delta g_n = \dfrac{1}{n+1}$

and at the Planck scale (n=0) the expression in eq. 9 gets the maximum value ($\Delta g_0 = 1$), which means that the light cone is smeared and the very concept of causality loses its meaning.

The emergence of causality above the Planck scale can be explained as follows. The causality degree is encoded in the program and this emerges as the network grows. Now, the growing of the network is an inherent property of the network as if it is encoded in the program itself.

Micro-causality can be then considered immediately above the Planck scale (already at scale $L_1 = 2L_P$), although it is not yet causality as we know it. The events of our causal set are not just points of a discrete spacetime, but the elements of a poset whose basic set is the set of qubits, whose order relation follows from the quantum entropy of the qubits, as illustrated in [1].

However, at the very Planck scale we cannot even speak of micro-causality, but just of proto-causality, in the sense that the future emergence of causality is encoded in the program. We believe that there exist a close relation between proto-causality and proto-consciousness [12] at the Planck scale.

## 4. Quantum computational aspects of spacetime foam

It was recently shown by Jack Ng [5], that spacetime foam provides limits to quantum computation. Here we investigate the quantum computational aspects of spacetime foam in the context of the QGN model.

We identify the node "n" as the quantum fluctuation of the metric on the $n^{nt}$ slice, that is, $\Delta g_n$ in eq. 9.

The energy of node "n" is then the energy [10] of the $n^{nt}$ quantum fluctuation of the metric:

10) $\quad E_n = \dfrac{E_P}{n+1} \qquad$ (n=0,1,2..)

where $E_P \approx 10^{19} GeV$ is the Planck energy.

From eq.10, we see that the energy of node "n" is the Planck energy divided by the number n+1 of virtual Planckian black holes (connecting links of the kind: $L_{-1,0}, L_{0,1}, L_{1,2}, ....L_{n-1,n}$ ). In particular, node "0" has energy $E_0 = E_P$.

In fact, due to the Planckian spacetime slicingin in eq. 4, a virtual black hole is created and annihilated (in between slice n and slice n+1), in the lapse of time $\Delta t = t_P$, with energy $\Delta E = E_P$. In fact, as it is well known, in this case the Heisenberg time-energy uncertainty relation is saturated:

11) $\quad \Delta t \Delta E = t_P E_P = \hbar$

Also, it should be noticed that the time-energy uncertainty relation is saturated at each time step:

12) $\quad t_n E_n = t_P E_P = \hbar$

By the use of eq. 9, eq. 10 can be rewritten as:

13) $\quad E_n = E_P \Delta g_n$

Or, by the use of eqs. 9 and 2, eq. 10 can be rewritten as:

14) $\quad E_n = E_P / \sqrt{I}$

Eq. 13 shows that the energy of the nodes is proportional to the quantum fluctuations of the metric. On the other hand, eq. 14 shows that the energy of the nodes is inversely proportional to (the square root of ) quantum information. This means that the oldest is a node, the lowest is its energy. The energy of the nodes decreases as the network grows.

This fact is due to the increase of the number of virtual black holes.
Moreover, from eq. 1 and eq. 9, it follows:

15) $\quad \Delta g_n = \sqrt{\Lambda}\, l_P$

As we have seen, both the quantum fluctuations of the metric and the cosmological constant are related to quantum information I. The relation between the cosmological constant and quantum information is given in eq. 1. The relation between quantum information and the quantum fluctuations of the metric is:

16) $\quad \Delta g_n = \dfrac{1}{\sqrt{I}}$

We define "quantum foam" the ensemble of the quantum fluctuations of the metric (the nodes) and the virtual black holes (connecting links $L_{-1,0}, L_{0,1}, L_{1,2}, ....L_{n-1,n}...$ ).

From the above arguments, it follows that the quantum foam is the seed of quantum computating spacetime.

What we call here "quantum foam" is based on the definition of "spacetime foam" due to Wheeler" [6], where quantum fluctuations of the metric induce fluctuations of the topology. Moreover, we enclose the quantum fluctuations of the vacuum (virtual black holes) in the definition of the "quantum foam" because, as we see from eq. 15, the two kinds of fluctuations are strictly related to each other in this context. Hawking [13] also considered virtual black holes as constituents of space-time foam. He showed that the topology of virtual black holes is:

17) $\quad S^2 \times S^2$

(the 4-dimensional spacetime beeing then a simply connected manifold). It should be noticed that these "bubbles" are not solutions of Einstein's field equations.

However, in our context at least, this is not the whole story. In fact, at node "0" one Euclidean Schwarzschild black hole of Planck size comes into existence, and it is not virtual, as it will give rise [10] to a quantum de Sitter spacetime. Its topology is:

18) $\quad R^1_{(space)} \times S^1_{(time)} \times S^2$.

It is clear that the factor $S^1$ in eq. 18, makes spacetime at the Planck scale a multiply connected manifold. Moreover, it should be noticed that the factor $S^1$ in eq. 18 is the periodic (imaginary) time: one more argument in favour of the acasual structure of spacetime at the Planck scale. Imaginary time is a important feature of the baby universes [14] scenario as well, which also prevents "travelling" from one universe to another.

This micro-black hole at node "0" has an horizon area of one pixel (one unit of Planck area) which, by the quantum holographic principle [1], encodes one unit of quantum information (one qubit). Our claim is that, virtual black holes carry virtual information which is transformed into real information by the quantum fluctuations of the metric (the logic gates). In the tranformation from virtual to real information, the topology itself must change, from eq. 17 to eq. 18.

Spacetime foam invokes the existence of a minimum length scale [15], which is the Planck length. The Planck length is the length scale of quantum gravity. This is made possible by the fact that general relativity is not a scale-invariant theory.

Given the Planck length as the minimum length scale, the proper distance between two events will never decrease beyond Planck length, and the uncertainty relation holds:

19) $\quad \Delta x \geq l_P$

Moreover, in our case, an analogous uncertainty relation holds for time intervals:

20) $\quad \Delta t \geq t_P$

as we are dealing with time quantized in Planck time units.

Then we get:

21) $\quad \Delta x \Delta t \geq l_P t_P$

For quantized spacetime in Planck units, it holds:

22) $\quad \Delta x = x_m - x_n = (m-n) l_P$
$\quad\quad \Delta t = t_m - t_n = (m-n) t_P$

and eq. 21 becomes:

23) $\quad \Delta x \Delta t = (m-n)^2 l_P t_P \geq l_P t_P$

The bound is saturated for m=n+1

Notice that we can also write:

24) $\quad x_n = \Delta g_n^{-1} l_P = \sqrt{I} l_P \; ; \quad t_n = \Delta g_n^{-1} t_P = \sqrt{I} t_P$

then, from eq. 24, it follows:

25) $\quad x_n t_n = \hbar \Delta g^{-2} = \hbar I$

where I is the quantum information given in eq. 2.

Thus, the quantum information encoded in quantum spacetime, is a consequence of the foamy structure of quantum spacetime itself.

Notice that for n=0, eq. 25 gives $x_0 t_0 = \hbar$, with I=1 (at the very Planck scale, quantum information encoded in the foam is one qubit).

## 5. The QGN as the ultimate Internet

The speed of computation $v$ of a system of average energy E, is bounded as [3]:

26) $\quad v \leq 2E/\pi\hbar$

Moreover, entropy limits the memory space $I$ [4] as the maximum entropy is given by:

27) $\quad S_{Max} = k_B I \ln 2$

where $k_B = 1.38 \times 10^{-23} J/K$ is the Boltzmann's constant

As Seth Lloyd showed [4], these bounds are saturated by black holes, which can then be regarded as ultimate computers.

In what follows, however, we will use the notations in [5], which are are more suitable to our purpose.

As Ng showed [5], the bounds on speed of computation $v$ and information $I$ can be reformulated respectively as:

28) $\quad v^2 \leq \dfrac{P}{\hbar}$

29) $\quad I \leq \dfrac{\hbar}{Pt_P^2}$

where P is the mean imput power.

Eq. 28 and eq. 29 lead to a simultaneous bound [5] on the information $I$ and the speed of computation $v$:

30) $\quad Iv^2 \leq \dfrac{1}{t_P^2}$

It is easily shown that the speed of computation $v_n$ and the quantum information $I_n$ of the QGN, saturate the three bounds above at each time step $t_n$:

31) $\quad v_n^2 = \dfrac{P_n}{\hbar} = \dfrac{E_n}{\hbar t_n} = \dfrac{\frac{E_P}{n+1}}{\hbar(n+1)t_P} = \dfrac{1}{(n+1)^2 t_P^2} = \dfrac{1}{t_n^2}$

32) $\quad I_n = \dfrac{\hbar}{P_n t_P^2} = \dfrac{\hbar}{\frac{E_n}{t_n}t_P^2} = \dfrac{\hbar}{\frac{E_P}{n+1}t_P^2} = (n+1)^2$

$\hspace{7em}\dfrac{}{(n+1)t_P}$

33) $\quad I_n v_n^2 = \dfrac{1}{t_P^2}$

As we have seen, at each time step, the QGN saturates the quantum limits to computation. Then, it can be regarded as a ultimate Internet, in the same sense that black holes can be regarded as ultimate computers.

Moreover, following Ng [5], we define the number of operations per time unit:

34) $\quad \bar{v} = Iv$

Eq. 34 becomes, in the context of the QGN:

35) $\bar{v}_n = I_n v_n$

Moreover, it holds:

36) $\bar{v}_n / \bar{v}_0 = n + 1 = \sqrt{I}$

where $\bar{v}_0$ is the number of operations per qubit at n=0:

37) $\bar{v}_0 = I_0 v_0;\quad v_0 = t_P^{-1} = 10^{43}\sec^{-1};\quad I_0 = 1$

From eq. 37, we get: $\bar{v}_0 = 10^{43}\sec^{-1}$.

From eq. 26, it follows that, if the average amount of energy is the Planck energy, $E_P \approx 10^{19} GeV \approx 10^9 J$, then, a Planckian black hole saturates the bound:

38) $v_P = \dfrac{2E_P}{\pi\hbar} \approx \dfrac{10^9 J}{10^{-34} J\sec} \approx 10^{43}\sec^{-1}$

This is the maximum speed of computation available in nature.

Also, from the simultaneous bound in eq. 30, it follows that the memory space of a Planckian black hole is one qubit:

39) $I_P = \dfrac{t_P^{-2}}{v_P^2} = 1$

Then, the qubit, the unit of quantum information, is the minimum amount of information available in nature.

This result is in agreement with the quantum holographic principle [1], which states that a Planckian black hole, having a horizon area of one pixel, can encode only one qubit.

By comparing eq. 37 with eqs. 38 and 39, it becomes evident that the node "0" is in fact a Planckian black hole.

Finally, it should be noticed that the information $I$, the speed of computation $v_n$ and the number of operations per time unit $\bar{v}_n$, are all related to the quantum fluctuations of the metric $\Delta g_n$:

40) $I = \Delta g_n^{-2};\quad v_n = \Delta g_n t_P^{-1};\quad \bar{v}_n = \Delta g_n^{-1} t_P^{-1}$

## 6.  The QGN and the problem of $\Lambda$

We calculate the present value of the quantized cosmological constant by the use of eq.1, with $n_{now} = 9 \times 10^{60}$, and we get:

41) $\Lambda_{now} = 1.25 \times 10^{-52} m^{-2}$

Also, we obtain:

42) $\Omega_\Lambda = \dfrac{\Lambda c^2}{3H_0^2} \approx \dfrac{1}{3}\Lambda c^2 t_{now}^2$

Where

43) $t_{now} \approx H_0^{-1} \approx 3 \times 10^{17} h^{-1}\sec$

$H_0$ beeing the Hubble constant and h a dimensionless parameter in the range: $0.58 \leq h \leq 0.72$. By choosing h=0.65, we get:

44) $\Omega_\Lambda \approx 0.7$

From the relation: $\Omega_\Lambda = \frac{\rho_\Lambda}{\rho_c}$, where $\rho_\Lambda$ is the vacuum energy density and $\rho_c$ is the critical density, it follows:

45) $\quad \rho_\Lambda \approx 0.7 \rho_c$

quite in agreement with the Type Ia SN observation data [8].
From this result, the connecting links (virtual states) of the QGN look like to be still active. But the value of the total entropy is too big, for $n_{now} = 10^{60}$: it would be $S_{now} = 10^{120} \ln 2$, indeed a huge amount of entropy. We believe that the at some earlier time, the QGN decohered. In this scenario, the free links were not activated anymore by the nodes, since the decoherence time. We can visualize the QGN after decoherence as a regular lattice, the connected part of the QGN itself.
From eq. 10, we have:

46) $\quad E_{now}^{NODES} = \frac{E_P}{n_{now}} = 10^{-41} GeV \approx 10^{-51} J$

One would expect that at present, such a weak energy of the nodes would prevent quantum computation. But this is not true: for $n \to \infty$, quantum information I will grow to infinity as $n^2$, while the energy of nodes will decrease to zero as $1/n$.
In fact, a low energy just reflects in a low speed of computation $\nu$ but not in a low amount of information.
From eq. 26 and eq. 46 it follows:

47) $\quad \nu_{now} \cong 10^{-17} \sec^{-1}$.

It should be noticed that in our case, $\nu_{now}^{-1}$ is the age of the universe: $\nu_{now}^{-1} \approx 10^{17}$ sec. Then, from the simultaneous bound in eq. 30, we get: $I_{now} \leq 10^{120}$, where in fact the bound is saturated because of the previous arguments. We would get in fact a huge total entropy $S_{now} = 10^{120} \ln 2$ if the machinery did not stop at some earlier time.

## 7.  Self-decoherence: a clue

The QGN can be theoretically divided into two sub-systems: the connected part, made of connecting links and nodes, (quantum fluctuations of the vacuum and quantum fluctuations of the metric respectively) and the disconnected part: the free links (qubits). The connected part can be thought as the "environment", while the disconnected part can be thought as the quantum state.
The virtual black holes in between spacetime slices, together with the quantum fluctuations of the metric on the slices, constitute spacetime foam.
Decoherence, here, can only be caused by spacetime foam. In general, if we consider a quantum process in foamy spacetime, this fuzzyness will induce decoherence of the wave function. This is a kind of back reaction of spacetime on the process [16]. In this particular case, the N qubits will decohere because of spacetime fuzzyness. The fact that virtual black holes can induce decoherence was suggested by Hawking [13]. However, as the quantum foam (virtual black holes and quantum fluctuations of the metric) resides in the connected subgraph, which is part of the QGN, the QGN as a whole undergoes self-decoherence. Although the details of the mechanism of self-decoherence are not yet properly understood, we can get a clue by considering entropy arguments, as it was already pointed out in [1] and [7].
The quantum entropy of N=I qubits is: S=Iln2.

Thus, if the QGN did never decohere, now the maximum entropy would be
$S = 10^{120} \ln 2$ ( with $n_{now} = 10^{60}$ ).
To get the actual entropy, one should compute it as:
$S_{now} = 10^{120} \ln 2 / S_{decoher} = 10^{120} / I_{Max}$
If one agrees with Penrose who claims [17] that the entropy now should be of order $10^{101}$, this corresponds to the maximum amount of quantum information at the moment of decoherence:

48) $\quad I_{Max} = 10^{19} \quad (n_{cr} \approx 10^9)$

where $n_{cr}$ stands for the critical number of nodes which are needed to process the maximum quantum information, $I_{Max}$.

From eq. 4, it follows that the early quantum computational universe decohered at
49) $\quad t_{decoh} \approx 10^{-34}$ sec.

Moreover, from eq. 6, we find that the mean energy at the moment of decoherence ($n = 10^9$) is:
50) $\quad E_{decoh} \approx 10^{10} GeV \approx 1J$

corresponding to a rest mass $m_{decoh} \approx 10^{-13} g$

From eq. 35, we get, for $n = 10^9$:
51) $\quad \bar{v}_{decoh} = I_{max} t_{decoh}^{-1} \approx 10^{53} \sec^{-1}$

Finally, it should be noticed that, as the QGN describes the early inflationary universe, the decoherence time corresponds to the time of the end of inflation, and the decoherence energy corresponds to the reheating energy.

## 8.     The QGN and the "information loss" puzzle

As we have seen above, the quantum logic gate at node "0" tranforms one virtual qubit into a real one. This is because node "0" is the zeroth quantum fluctuation of the metric, which can be identified [10] with one Planckian Euclidean Schwarzschild black hole. The latter, has a horizon area of one pixel, which, by the quantum holographic principle [1] encodes one qubit of information.
This fact might be relevant when considering the "information loss" puzzle in black hole evaporation (for a review on this subject see for example [9]).
Hawking's radiation is the thermal emission from a classical black hole [18]. Hawking used semiclassical approximations (quantum effects in a curved spacetime), which lead him to argue that the black hole would ultimately evaporate completely, and all the information collapsed into the black hole would be lost [19]. In more precise terms, a pure quantum state that collapsed into a black hole, would come out as a mixed state of thermal radiation. This process is described by a superscattering operator which transforms the initial density matrix in the final one in a nonunitary way. Many attempts have been made to overcome the information loss problem: one of the most famous is the proposal of the remnants.
The belief, is that the semiclassical arguments of black holes evaporation might fail at the Planck scale. When the black hole reaches the Planck mass, strong quantum gravity effects might stop the evaporation process. In this case, there would be a remnant, which should store all the information collapsed in the original black hole. However, if the remnant is a black hole itself (of Planck size), this cannot happen. In fact, as we have seen above, the horizon area of a Planckian black hole can encode only one qubit of information. Certainly, the remnant is not a virtual black hole. The alternative is that the remnant Planckian black hole gives rise to a QGN, (a quantum de Sitter universe) to regain the "lost" number of bits. This idea is similar to the original one of Dyson [20], that the black hole disappears completely, but one or more new universes branch off and carry away the information.

The QGN programming a new universe provides a number N of qubits which is equal to the number of bits lost inside a black hole in a older universe. This number N determines the information entropy $S(T_N) = N \ln 2$ of the new universe at time $T_N = \sqrt{N} t_P$ (N=1,2,3…). The entropy at a later time $T_{N'} > T_N$, which would be the maximum entropy at that time, that is, $S_{N'} = N' \ln 2$, is in effect reduced to the actual entropy at that time by a factor N:

52) $$S(T_{N'}) = \frac{N'}{N} \ln 2$$

One could argue then that only N qubits can be produced by the QGN, or, alternatively, that after the production of N qubits, even if the maximum number was not programmed, the QGN self-decoheres, as illustraded in the previous section. However, the two pictures are equivalent. In fact, the structure of spacetime foam causing decoherence is part of the program.

Eventually, we are faced with a cosmic system of universes, each one beeing a QGN (a ultimate Internet). Information is conserved in the cosmic system, because of the enlargement of the Hilbert space, and this in agreement with the ideas expressed by Hitchcock in a recent paper [21].

It should be noticed that what we called the "unphysical time" $t_{-1} = 0$ (corresponding to a singularity in the classical theory) is unphysical for the new born universe, but not for the mother universe. In fact, in the mother universe, $t_{-1}$ is the latest instant of evaporation of the black hole which will originate the child universe. The fact that there is this "leap" from physical time to unphysical time, in the passage from one universe to another, means that the two universes are not causally related. This is due to the fact that the causality degree is zero at the Planck scale (at $t_0$) and undefined beyond it (for example at $t_{-1}$) as we have seen in Section 3.

In this scenario, the following arguments hold:
i) The more black holes an old universe has got, the more new universes it can produce.
ii) Information is lost in black holes of old universes to lower the value of the entropy of young universes, and reduction of entropy is a survival tendency.
iii) New born universes originating from a big information loss (a large N), will enflate more than new born universes originating from a small information loss. In fact, from eq. 30, we see that for small $N \equiv I$, inflation will be short (small $\nu^{-1}$) and from eq. 34, we also see that the speed of computation $\bar{\nu}$ (number of computations per time unit) will be small.

All this is very much on line with the idea expressed by Smolin in his book "The life of the cosmos" [22], that the cosmos evolves according to a principle similar to that of natural selection.

This system of ultimate Internets can be regardeds as a self-organizing system. In fact, the new born universe is produced in such a quantum state leading, after decoherence, to a classical universe with reduced maximum entropy, for which evolution is possible. Similarly, in a recent paper by Jaroszkiewicz [23], the universe was described as a autonomous, self-organizing quantum automaton.

It does not seem necessary for the whole cosmic system to have a "beginning" in the usual sense of the Big Bang (and inflation).

But there is a "beginning" for each universe. The beginning of a new universe is when it starts to get its own identity, when the umbilical cord with the mother universe breaks. This happens with a "leap" from unphysical time (which was the last physical instant of the black hole evaporation in the older universe) to physical time (when information starts to be associated with time in the new born universe). That is what we call the Planck time in our universe. The Planck length must be unpassable for the same reason: there is no information associated with a smaller scale than the Planck scale (some work is in progress [12] on this topic).

There is no fundamental reason that the number of qubits at which our inflationary universe self-decohered should be $10^{19}$. This was just the number of bits lost in the evaporation of a black hole in our mother universe. Nevertheless, this number is important to us. In fact, because of it, the present entropy of our universe is smaller than the maximum possible entropy by a factor $10^{19}$. We believe that this number is responsible for the particular evolution of our universe.

In the same way, all the other universes got their own programs for evolution. We wish to make a distinction here between oblique universes and parallel universes. We name oblique universes those which were generated by black holes evaporation with different information loss, and which will have totally different evolutions. We are not even able to figure out a different evolution from that of our universe, because we ourselves are a product of a specific evolution, which was encoded in a specific program. Thus, we are not only causally unrelated, but also logically unrelated, to oblique universes.

Instead, we call parallel universes those which were generated from black holes evaporation with the same information loss, and which will have similar evolutions. We are able to understand the evolution of universes which are parallel to ours, but still, we are not in causal relation with them.

Thus, there is no communication among ultimate Internets.

It should be noticed that here the definition of "parallel universes" means something different from the usual definition used in the many worlds interpretation of quantum mechanics [24] and in the multiverse [25]. In fact, the multiverse is the domain of parallel universes, in the sense that a universe exists for every physical possibility (outcome) of a quantum event. In our model, each QGN is a multiverse itself (actually, Deutsch descibes the multiverse as a quantum computer). However, in our case, that is not the end of the story. Outside our particular QGN, there are others QGNs which can be parallel to ours, in the sense that they can have an evolution similar to that of our universe, maybe leading to humankind as well. In fact our model, is similar to the multiverse scenario for each QGN, but it is also reminiscent of the baby universes picture [14] when all possible QGNs are taken into account.

In conclusion, the maximum quantum information N of a QGN, will influence the future evolution of that universe (and of its possible inhabitants).

## 9. Concluding remarks

As we have seen, at its very beginning, a new universe is in no casual relation with anything. There is only one qubit of information, the universe itself, and any causal relation with the mother universe was erased by the "leap" from physical to unphysical time.

Causality emerges with the growth of the QGN. In this context, then, physical reality is due to information: there is no existence without information. And causality is an emergent feature: there can be no causality without evolution, although proto-causality is encoded in the program already at the Planck scale.

In [26], we suggested that, when in our universe the $10^{19}$ qubits decohered (end of inflation), the universe itself underwent a kind of conscious experience. This idea was based mostly on the fact that the number of tubulins/qubits in our brain is only one order of magnitude less, and that decoherence of tubulins in the brain originates a conscious experience [27]. However, the idea of cosmic consciousness can be applied also to the other universes of the cosmic system. The number of qubits involved in the decoherence of a QGN, determines also the evolution of future conscious minds in the universe associated with that QGN.

In this scenario, consciousness appears to be strictly related to information, causality, and evolution. Consciousness must be an emergent feature, although its roots are in quantum spacetime [17]. Proto-consciousness, like proto-causality, is encoded in the QGN at the Planck scale [12].

Quantum gravity seems to be more complex than just the unification of quantum mechanics and general relativity, as apparently it involves quantum computing and quantum communication.

While completing the references list of this paper, we became aware of two very recent and interesting papers.

The first one is a paper by Finkelstein and coworkers [28], where a system of elementary operations is proposed, generating all physical operations, which can be regarded as performed by a cosmic quantum computer.

The second one, is a paper by Lloyd [29] (whose outcoming was pre-announced to us by the author some time ago), where the memory space and the number of elementary operations of the universe are quantified. The values that Lloyd finds, are quite in agreement with ours. The maximum possible number of bits which the universe registered since the Big Bang until now beeing $10^{120}$, equal to the number of elementary operations, which is calculated by the use of the relation: $(t/t_P)^2$, where in our case t is the quantized time $t_n$ and the relation above gives explicitly: $(t_n/t_P)^2 = (n+1)^2 \equiv N$. The main difference is that Lloyd does not consider decoherence at $N = 10^{19}$, while we do.

## Aknowledgments

I am grateful to Sisir Roy, Scott Hitchcock, Seth Lloyd, Mitchell Porter and Luis Garay for useful discussions.

## References:

[1]  P. A. Zizzi, "Holography, Quantum Geometry and Quantum Information Theory", gr-qc/9907063, Entropy 2 (2000) 39.

[2]  G. ' t Hooft, "Dimensional reduction in Quantum Gravity", gr-qc/9310026;
G. ' t Hooft, "The Holographic Principle", hep-th/0003004;
L. Susskind, "The World as a Hologram", hep-th/9409089.

[3]  N. Margolous, L. B. Levitin, Physica D120(1988) 188.

[4]  S. Lloyd, "Ultimate physical limits to computation", quant-ph/9908043;
Nature 406 (2000) 1047.

[5]  Y. J. Ng, " Clocks, computers, black holes, spacetime foam and holographic principle", hep-th/0010234;
Y. J. Ng, ""From computation to black holes and space-time foam", gr-qc/0006105.

[6]  J. A. Wheeler, "Geometrodynamics", Academic Press, New York (1962).

[7]  P. A. Zizzi, "The Early Universe as a Quantum Growing Network", gr-qc/0103002, in Proceedings of IQSA Fifth Conference, March 31-April 5, 2001, Cesena-Cesenatico, Italy.

[8]  S. Perlmutter et al., Nature 391 (1998) 51;
B. P. Schmidt et al., Astrophys. J. 507 (1998) 46.

[9]  T. Banks, "Lectures on Black Holes and Information Loss", hep-th/9412131.

[10] P. A. Zizzi, "Quantum Foam and de Sitter-like Universe", hep-th/9808180, Int. J. Theor. Phys. 38 (1999) 2333.

[11] Albert-Laszlo Barabasi and Reka Albert, "Emergence of Scaling in Random Networks", cond-mat/9910332;
S. N. Dorogowtsev, J. F. F. Mendes, and A. N. Samukhin, "WWW and Internet models from 1955 till our days and the "popularity is attractive" principle", cond-mat/0009090.

[12] S. Hameroff and P. A. Zizzi, work in progress.

[13] S. W. Hawking, D. N. Page and C. N. Pope, "Quantum gravitational bubbles", Nucl. Phys. B 170 (1980) 283;
S. W. Hawking, "Virtual black holes", Phys. Rev. D 53 (1996) 3099.

[14] S. W. Hawking, "Black Holes and Baby Universes and other Essays", Bantam, New York (1993).

[15] Luis J. Garay, "Quantum gravity and minimum length", gr-qc/9403008.

[16] Sisir Roy, "Statistical Geometry and Application to Microphysics and Cosmology", Kluwer Academic Press, The Netherlands (1998).

[17] R. Penrose, "The Emperor 's New Mind", Oxford University Press (1989).

[18] S. W. Hawking, Comm. Math. Phys. 43 (1975)199.

[19] S. W. Hawking, Phys. Rev. D14 (1976) 2460.

[20] F. Dyson, Institute for Advanced Study preprint, 1976, unpublished.

[21] Scott M. Hitchcock, "Is there a Conservation of Information Law for the Universe?", gr-qc/0108010.

[22] Lee smolin, "The Life of the Cosmos", Oxford University Press, New York (1997).

[23] George Jaroszkiewicz, "The running of the universe and the quantum structure of time", quant-ph/0105013.

[24] H. Everett III, "Relative State Formulation of Quantum Mechanics", Rev. of Modern Phys. Vol. 29 (1957) 454.

[25] David Deutsch, "The structure of the multiverse", quant-ph/0104033.
[26] P. A. Zizzi, "Emergent Consciousness: From the Early Universe to our Mind", gr-qc/0007006.
[27] S. Hameroff and R. Penrose, "Orchestrated reduction of quantum coherence in brain microtubules: A model for consciousness", in: Toward a Science of Consciousness. The first Tucson Discussions and debates, MIT Press, Cambridge, MA (1996).
[28] J. Baugh, A. Galiautdinov, D. R. Finkelstein and H. Saller, " Elementary Quantum Operations", in Proceedings of IQSA Fifth Conference, March 31- April 5, 2001, Cesena-Cesenatico, Italy.
[29] Seth Lloyd, "Computational Capacity of the Universe", quant-ph/0110141.